\documentclass[onecolumn,showpacs,prd]{revtex4}
\usepackage{CJK}                     
\usepackage[dvips]{graphicx}
\usepackage{bm}                      
\usepackage{mathptmx}                
\usepackage{amssymb}  
\usepackage{dcolumn}
\usepackage{epsfig}
\usepackage{hyperref}
\usepackage{graphicx}
\usepackage{longtable}

\usepackage[dvips]{color}

\usepackage[normalem]{ulem}  

\newcommand{\NPA}[3]{Nucl.\ Phys.\ {\bf A#1}, #2 (#3)}

\newcommand{\PLB}[3]{Phys.\ Lett.\ B\ {\bf #1}, #2 (#3)}

\newcommand{\PRL}[3]{Phys.\ Rev.\ Lett.\ {\bf #1}, #2 (#3)}

\newcommand{\PRC}[3]{Phys.\ Rev.\ C\ {\bf #1}, #2 (#3)}
\newcommand{\PRD}[3]{Phys.\ Rev.\ D\ {\bf #1}, #2 (#3)}
\newcommand{\JPG}[3]{J.\ Phys.\ G\ {\bf #1}, #2 (#3)}

\begin{document}

\title{Magnetized strange quark matter in a quasiparticle description}
\begin{CJK*}{GB}{gbsn}

\author{
Xin-Jian Wen,$^{\mathrm{1}}$\footnote{E-mail address:
wenxj@sxu.edu.cn} Shou-Zheng Su,$^{\mathrm{1}}$ Dong-Hong
Yang,$^{\mathrm{1}}$ and Guang-Xiong Peng $^{2,3}$\footnote{E-mail
address: gxpeng@gucas.ac.cn} }

\affiliation{ $^1$Department of Physics and Institute of Theoretical
Physics, Shanxi University, Taiyuan 030006, China\\
$^2$College of Physics, Graduate University of Chinese
Academy of Sciences, Beijing 100049, China\\
$^3$ Theor.\ Physics Center for Science Facilities, CAS Institute of
High Energy Physics, Beijing 100049, China
              }
\date{\today}

\begin{abstract}
The quasiparticle model is extended to investigate the properties of
strange quark matter in a strong magnetic field at finite densities.
For the density-dependent quark mass, self-consistent thermodynamic
treatment is obtained with an additional effective bag parameter,
which depends not only on the density but also on the magnetic field
strength. The magnetic field makes strange quark matter more stable
energetically when the magnetic field strength is less than a
critical value of the order $10^7$ Gauss depending on the QCD scale
$\Lambda$. Instead of being a monotonic function of the density for
the QCD scale parameter $\Lambda>126$~MeV, the effective bag
function has a maximum near $0.3\sim 0.4$~fm$^{-3}$. The influence
of the magnetic field and the QCD scale parameter on the stiffness
of the equation of state of the magnetized strange quark matter and
the possible maximum mass of strange stars are discussed.

\end{abstract}

\maketitle
\end{CJK*}

\section{Introduction}
\label{sec:intro} Since strange quark matter (SQM) was speculated by
Witten as the possible true ground state of strong interaction
matter \cite{witten}, the properties of SQM in bulk, as well as in
finite size, the so called strangelets, have been extensively
studied in the past decades
\cite{Farhi84,Buballa96,Madsen00,Dicus08}. The new form of matter is
possibly produced by terrestrial relativistic heavy-ion collision
experiments \cite{Heinz01} or exists in the interior of compact
stars \cite{Alcock86}. It was found that the stability of SQM is
strongly affected in a strong magnetic field \cite{Chakra1996}. The
large magnetic fields in nature are normally associated with
astrophysical objects, where the density is much higher than the
nuclear saturation. The typical strength could be of the order $\sim
10^{12}$~G on the surface of pulsars~\cite{Dong1991}. Some magnetars
can have even larger magnetic fields, reaching the surface value as
large as $10^{14}\sim 10^{15}$ G \cite{pacz}. In the interior of
compact stars, the maximum possible magnetic field strength is
estimated as high as $\sim 10^{18}$~G. The origin of the strong
magnetic fields can be understood in two ways. One is the
amplification of the relatively small magnetic field during the
star's collapse with magnetic flux conservation \cite{Tatsumi06}.
The other is the magnetohydrodynamic dynamo mechanism with large
magnetic fields generated by rotating plasma of a protoneutron star
\cite{Vink06}.

Because a strong magnetic field influences the single particle
spectrum while all quarks are charged, SQM in the inner part of a
compact star may show specific properties. Specially, for example,
the strong magnetic field leads to a more stable polarized strange
quark star (SQS)\cite{Bordbar11}. In heavy-ion collisions
experiments, the magnitude of a magnetic field plays an important
role in studying the deconfinement and chiral phase transitions. In
the LHC/CERN energy, it is possible to produce a field as large as
$5\times 10^{19}$~G \cite{Skokov09}.

With various phenomenological confinement models, many works on the
properties of magnetized SQM(MSQM) have been done by a lot of
researchers. Based on the conventional MIT bag model, quark matter
in a strong magnetic field was studied by Chakrabarty
\cite{Chakra1996}, and significant effect on the equation of state
had been found. Furthermore, the magnetized strangelets at finite
temperature was investigated by Felipe {\sl et. al.} in their recent
work \cite{Felipe,Felipe2011}. In Ref.~\cite{Mizher10}, the effect
of an external magnetic field on the chiral dynamics and confining
properties of SQM were discussed in the linear sigma model coupled
to the Polyakov loops. The special properties of MSQM were also
investigated with the Nambu-Jona-Lasinio (NJL) model
\cite{Ebert03,Frol10,Faya11,Avan11}. The MIT bag model, the
two-flavor NJL model, and the chiral sigma model had also been
compared in studying the MSQM \cite{Rabhi2011}.

In literature, the quasiparticle model, where the effective quark
mass varies with environment, was also successfully employed by many
authors to study the dense strange quark matter in the absence of an
external magnetic field~\cite{vija95,Schertler1997,peshier2000}. The
main advantage of the quasiparticle model is that it can explicitly
describe quark confinement and vacuum energy density for bulk
matter~\cite{Schertler1997} and strangelets~\cite{wen2010}. The aim
of this article is to extend the quark quasiparticle model to
studying the magnetized quark matter. We find a density- and
magnetic-field-dependent bag function. Accordingly, a
self-consistent thermodynamic treatment is obtained with the new
version of the bag function. The effect of a magnetic field on the
bag function and the stability of MSQM will be discussed. It is
found that the magnetic field makes SQM more stable when the
magnetic field strength is less than a critical value of the order
$10^7$~G depending on the QCD scale $\Lambda$.

This paper is organized as follows. In Sec. 2, we derive the
thermodynamic formulas in the quasiparticle model when the magnetic
field becomes rather important, and then demonstrate the effective
bag function for the case of both constant and running coupling,
respectively. In Sec. 3, the stability properties of MSQM, the
effective bag function, and the mass-radius relation of magnetized
quark stars are investigated, and discussions are shown about the
effect of the magnetic field and QCD scale parameter. The last
section is a short summary.
\section{thermodynamic treatment in a strong magnetic field
}\label{thermody} The important feature of the quasiparticle model
is the medium dependence of quark masses in describing QCD
nonperturbative properties. The quasiparticle quark mass is derived
at the zero-momentum limit of the dispersion relations from an
effective quark propagator by resuming one-loop self-energy diagrams
in the hard dense loop (HDL) approximation. In this paper, the
effective quark mass is adopted as
\cite{Schertler1997,Schertler1997jpg,Pisarski1989}
\begin{eqnarray}m_i(\mu_i)=\frac{m_{i0}}{2}+ \sqrt{\frac{m^2_{i0}}{4}+
\frac{g^2\mu_i^2}{6\pi^2}}\, ,\label{mass1}
\end{eqnarray}
where $m_{i0}$ and $\mu_i$ are, respectively, the quark current mass
and chemical potential of the quark flavor $i$. The constant $g$ is
the strong interaction coupling. One can also use a running coupling
constant $g(Q/\Lambda)$ in the equations of state of strange matter
instead of a constant $g$ \cite{shir1997}. In our recent work by
using phenomenological running coupling \cite{wen2010}, the quark
masses were demonstrated to decrease with increasing densities at a
proper region.

Here, we assume the $g$ value is in the range of $(0, 0.5)$, as done
in the previous work \cite{Schertler1997}. The current mass can be
neglected for up and down quarks, while the strange quark current
mass is taken to be $120$ MeV in the present calculations. Because
the vanishing current mass is assumed for up and down quarks,
Eq.~(\ref{mass1}) is reduced to the simple form
\begin{eqnarray}m_i=\frac{g \mu_i}{\sqrt{6}\pi}.
\end{eqnarray}

Instead of inserting the effective mass $m_i$ directly into the
Fermi gas expression, we will derive the expressions from the
self-consistency requirement of thermodynamics. The quasiparticle
contribution of the flavor $i$ to the total thermodynamic potential
density can be written as
\begin{eqnarray}
\Omega_i &=&
   -\frac{d_i T}{(2\pi)^3}
   \int_0^{\infty}
    \left\{
 \ln\left[1+e^{-(\epsilon_{i,p}-\mu_i)/T}\right]\right. \nonumber\\
 &&\phantom{ -\frac{d_i T}{(2\pi)^3}
   \int_0^\infty  \{} \left.
   +\ln\left[1+e^{-(\epsilon_{i,p}+\mu_i)/T}\right]
    \right\}
  \mbox{d}^3\vec{p},\label{omegaT}
\end{eqnarray}
 where $T$ is the system temperature and $d_i$ is the degeneracy factor
[$d_i= 3(\rm{color})$ for quarks and $d_i=1$ for electrons]. All the
thermodynamic quantities can be derived from the characteristic
function by obeying the self-consistent relation \cite{wen2009}.

To definitely describe the magnetic field of a compact star, we
assume a constant magnetic field ($B_{m,z}=B_m$) along the $z$ axis.
Due to the quantization of orbital motion of charged particles in
the presence of a strong magnetic field, known as Landau
diamagnetism, the single particle energy spectrum is \cite{Landau}
\begin{eqnarray}\varepsilon_i=\sqrt{p_z^2+m_i^2+e_iB_m(2n+s+1)},
\end{eqnarray}
where $p_z$ is the component of particle momentum along the
direction of the magnetic field $B_m$, $e_i$ is the absolute value
of the electronic charge (e.g., $e_i=2/3$ for the up quark and 1/3
for the down and strange quarks), $n=0,1,2,...$, are the principal
quantum numbers for the allowed Landau levels, and $s=\pm 1$ refers
to quark spin-up and -down state, respectively. For the sake of
convenience, we set $2\nu=2n+s+1$, where $\nu=0,1,2,...$. The single
particle energy then becomes~\cite{Chakra1996}
\begin{eqnarray}\varepsilon_i=\sqrt{p_z^2+m_i^2+2\nu e_iB_m}.
\end{eqnarray}
On application of the quantized energy levels, the integration over
$dp_xdp_y$ in Eq. (\ref{omegaT}) is replaced by the rule
\begin{eqnarray}\int_{-\infty}^{+\infty}
\int_{-\infty}^{+\infty}dp_xdp_y \rightarrow 2\pi e_iB_m\sum_{s=\pm
1} \sum_n .
\end{eqnarray}
Because there is the single degenerate state for $\nu=0$ and the
double degenerate state for $\nu\neq 0$, we assign the spin
degeneracy factor ($2-\delta_{\nu 0}$) to the index $\nu$ Landau
level. The thermodynamic potential density of Eq.(\ref{omegaT}) in
the presence of a strong field can thus be written as
\begin{eqnarray}\Omega_i(T,m_i,\mu_i)=-T\frac{d_ie_iB_m}{2\pi^2}\sum_{v=0} (2-\delta_{\nu
0}) \int_0^\infty \Big\{\ln[
{1+\exp(\frac{\mu_i-\varepsilon_i}{T})}]+\ln[
{1+\exp(\frac{-\mu_i-\varepsilon_i}{T})}] \Big\}dp_z.
\label{eq:omega}
\end{eqnarray}
At zero temperature, Eq.~(\ref{eq:omega}) is simplified to give
\begin{eqnarray}\Omega_i(m_i,\mu_i)&=&-\frac{d_ie_iB_m}{2\pi^2}\sum_{v=0}(2-\delta_{\nu
0})\int_0^{\sqrt{\mu_i^2-M_\nu^{(i)2}}} (\mu_i-\varepsilon_i)dp_z \nonumber\\
&=&-\frac{d_ie_iB_m}{2\pi^2}\sum_{v=0}^{\nu_{max}}(2-\delta_{\nu
0})\Big\{
\frac{1}{2}\mu_i\sqrt{\mu_i^2-M_\nu^{(i)2}}-\frac{1}{2}M_\nu^{(i)2}\ln(\frac{\mu_i+(\mu_i^2-{M_\nu^{(i)2}})^{1/2}}{M_\nu^{(i)}})
\Big\},\label{omega0}
\end{eqnarray}
where $M_\nu^{(i)}=\sqrt{m_i^2+2\nu e_i B_m}$ is the quark effective
mass in the presence of a magnetic field. In the case of zero
temperature, the upper limit $\nu_{max}$ of the summation index
$\nu$ can be understood from the positive value requirement on the
logarithm and square-root function in Eq. (\ref{omega0}). So we have
\begin{equation}\nu\leq \nu_{max}\equiv
\mathrm{int}[\frac{\mu_i^2-m_i^2}{2e_iB_m}],
\end{equation}where ``int" means the number before the decimal point.

Accordingly, the pressure $P$, the energy density $E$, and the free
energy density $F$ for SQM at zero temperature read~\cite{wen2005}

\begin{eqnarray}P&=&-\Omega_f -B^*,\label{press}\\
E &=&F= \Omega_f+\sum_i \mu_i n_i +B^*. \label{energy}
\end{eqnarray}
Here $\Omega_f=\sum_i\Omega_i$ is the free quasiparticle
contribution with the summation index going over all flavors
considered. The notation $B^*$ denotes the effective bag function
and it can be divided into two parts: $\mu_i$-dependent part and the
definite integral constant part, i.e., $B^*=\sum_i B_i(\mu_i)+B_0$
($i=u$, $d$, and $s$) where $B_0$ is similar to the conventional bag
constant and $B_i(\mu_i)$ is the chemical potential dependent
function to be determined.

The derivative of the thermodynamic potential density $\Omega_i$
with respect to the quark effective mass $m_i$ has an analytical
expression, i.e.,
\begin{eqnarray}
\frac{\partial \Omega_i}{\partial m_i}=\frac{\partial
\Omega_i}{\partial M_\nu^{(i)}}\frac{\partial M_\nu^{(i)}}{\partial
m_i}=\frac{d_ie_iB_m}{2\pi^2}\sum_{v=0}^{\nu_{max}}(2-\delta_{\nu
0})
m_i\ln[\frac{\mu_i+(\mu_i^2-M_\nu^{(i)2})^{1/2}}{M_\nu^{(i)}}].\label{eq:dOdm}
\end{eqnarray}

The quark particle number density of the component $i$ is given as
\begin{eqnarray}n_i=\frac{d_i
e_iB_m}{2\pi^2}\sum_{\nu=0}^{\nu_{max}}(2-\delta_{\nu
0})\sqrt{\mu_i^2-M_\nu^{(i)2}}. \label{eq:ni}
\end{eqnarray}

In the literature, there are three methods to construct a consistent
set of thermodynamical functions with the effective quark masses.
One is applied in the quark mass density-dependent model in
Refs.\cite{Chakra91,peng99}, where all thermodynamic quantities are
derived by direct explicit function and implicit function dependent
relations. The second is the treatment in the NJL model, where the
dynamical quark masses are solutions of the gap equation coupling
the quark condensates \cite{buba99,Avan11}. The energy and pressure
functions are modified accordingly. The third method is to get a
self-consistent thermodynamical treatment with an effective bag
constant to describe the residual interaction \cite{Romat03}. The
effective bag constant acts as a part of a modified pressure
function. Here, we employ the third method. The following
requirement is introduced and applied as in
Refs.~\cite{Goren1995,Schertler1997},
\begin{eqnarray}\left(\frac{\partial P}{\partial
m_i}\right)_{\mu_i}=0. \label{dp=0}
\end{eqnarray}
From a physical viewpoint, the constraint can make the formula of
particle number function consistent with standard statistical
mechanics. From Eqs. (\ref{press}) and (\ref{energy}), it can be
understood that the effective bag constant leads an additional term
in the modification in the energy and pressure functions.

Considering Eq.(\ref{dp=0}), we have the vacuum energy density
$B_i(\mu_i)$ through the following differential equation
\begin{eqnarray}\frac{\mathrm{d} B_i(\mu_i)}{\mathrm{d} \mu_i}\frac{\mathrm{d} \mu_i}{\mathrm{d} m_i}=-\frac{\partial \Omega_f}{\partial
m_i}.\label{eq:dBdm}
\end{eqnarray}
If we assume the vanishing current quark mass, one can integrate
Eq.~(\ref{eq:dBdm}) under the condition $B_i(\mu_i=0)=0$ and have
\begin{eqnarray}\label{Bexp}
B_i(\mu_i)&=&-\int^{\mu_i}_0 \left.\frac{\partial \Omega_f}{\partial
m^*_i}\right|_{T=0,\mu_i}\frac{\mathrm{d} m_i}{\mathrm{d}\mu_i} \mathrm{d}\mu_i \nonumber \\
&=&-\frac{d_ie_iB_m}{2\pi^2} \sum_{\nu=0}^{\nu_{max}} (2-\delta_{\nu
0}) \int_{\mu_i^c}^{\mu_i}\alpha^2
\mu_i\ln(\frac{\mu_i+\sqrt{\mu_i^2-M_\nu^{(i)2}}}{M_\nu^{(i)}})
d\mu_i,
\end{eqnarray}
where the lower limit of the integration over $\mu_i$ is different
from that in Ref. \cite{Schertler1997}. Its critical value $\mu_i^c$
should satisfy
\begin{equation}\label{muceq}\mu_i^{c2}-m_i^2-2\nu e_iB_m\geq0
\end{equation}

To reflect the asymptotic freedom of QCD, the calculation must be
changed by including the running coupling constant. The approximate
expression for the running quantity $g(\mu)$ reads~\cite{Patra1996},
\begin{equation}\label{g_run}g^2(T=0,\mu)=\frac{48 \pi^2}{29}\left[\ln(\frac{0.8
\mu^2}{\Lambda^2})\right]^{-1}, \label{rung}
\end{equation}
where $\Lambda$ is the QCD scale parameter, the only free parameter
in the theory determined by experiments. The magnitude of $\Lambda$
controls the rate at which QCD coupling constant runs as a function
of exchanged momentum $Q^2$ (see Ref.~\cite{shir1997}). After
applying the running coupling constant (\ref{rung}), the effective
bag function in Eq.~(\ref{Bexp}) is changed into
\begin{eqnarray}\label{Bexp1}
B_i(\mu_i)&=&-\int^{\mu_i}_0 \left.\frac{\partial \Omega_f}{\partial
m^*_i}\right|_{T=0,\mu_i}\frac{\mathrm{d} m_i}{\mathrm{d}\mu_i} \mathrm{d}\mu_i \nonumber \\
&=&-\frac{d_ie_iB_m}{2\pi^2} \sum_{\nu=0}^{\nu_{max}}(2-\delta_{\nu
0}) \int_{\mu_i^c}^{\mu_i}
m_i\ln(\frac{\mu_i+\sqrt{\mu_i^2-M_\nu^{(i)2}}}{M_\nu^{(i)}})\frac{\mathrm{d}
m_i(\mu_i,g(\mu_i))}{\mathrm{d} \mu_i} d\mu_i,
\end{eqnarray}where the lower limit of the integration $\mu_i^c$ satisfies
$B_i(\mu_i^c=0)$. Differently from the constant coupling case, the
critical value $\mu_i^c$ can be obtained by inserting the running
coupling constant in Eq.~(\ref{g_run}) into the condition
(\ref{muceq}). The value of $\mu_i^c$ depends not only on the
chemical potential of quarks but also on the Landau energy level.

\section{properties of magnetized Strange quark matter}
In this section, the properties of MSQM are studied with the new
version of the quasiparticle model in the presence of a strong
magnetic field. We will investigate the properties with a density-
and magnetic-field-dependent bag function. Then, we discuss the
effect of the QCD scale parameter and the strong magnetic field on
the effective bag function and strange quark stars.

\subsection{The stability property of bulk magnetized SQM}

As is usually done, the SQM is treated as a mixture of $u$-, $d$-,
$s$- quarks and electrons with neutrinos entering and leaving the
system freely. To obtain the equations of state (EoS) of magnetized
SQM, a set of equilibrium conditions--the weak equilibrium, baryon
number conservation, and electric charge neutrality--should be
considered by the following relations
\cite{Chakra1996,band1997,Singh02,Gonz08,Felipe}:
\begin{eqnarray}&&\mu_u+\mu_e=\mu_d=\mu_s,\label{eq:chemi}\\
&&n_u+n_d+n_s=3n_\mathrm{B},\label{eq:nb}\\
&&\frac{2}{3}n_u-\frac{1}{3}n_d-\frac{1}{3}n_s-n_e=0.\label{eq:charge}
\end{eqnarray}

Equation~(\ref{eq:chemi}) is the chemical equilibrium condition
maintained by the weak-interaction processes such as $s+u\rightarrow
u+d$ and $s\rightarrow u+e+\bar{\nu}_e$ etc., Eq.~(\ref{eq:nb}) is
from the definition of the baryon number density $n_\mathrm{B}$, and
Eq.~(\ref{eq:charge}) is the charge neutrality condition. For a
given baryon number density $n_\mathrm{B}$, we can obtain the four
chemical potentials $\mu_u$, $\mu_d$, $\mu_s$, and $\mu_e$ by
solving the four equations in Eq.
(\ref{eq:chemi})-(\ref{eq:charge}). Other thermodynamic quantities,
such as the energy density and pressure, can then be calculated from
the formulae derived in the previous section. A little difference is
that the Maxwell contribution has been included in our numerical
calculations, i.e., the quasiparticle contribution $\Omega_f$ is
replaced by \cite{noron07,Mene09,Ferrer10}
\begin{eqnarray}\Omega=\Omega_f+\frac{B_m^2}{2},
\end{eqnarray}where the second term is the pure Maxwell contribution
of the magnetic field itself.

\begin{figure}[htb]
\centering
\includegraphics[width=7cm,height=7cm]{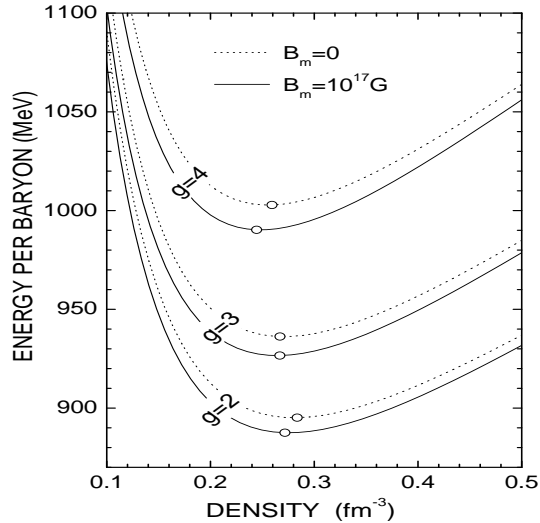}
\caption{The energy per baryon versus the density at fixed coupling
constant $g=2,3,4$ for magnetic field strength B$_m=10^{17}$G.
Compared with the nonmagnetized strange quark matter (the dotted
curves with $B_m=0$), the magnetized case has a lower energy per
baryon.
        }
\label{En}
\end{figure}

In Fig.~\ref{En}, the energy per baryon of MSQM is shown as
functions of the density for several $g$ values. For comparison
purposes, we have also plotted the previous results in
Ref.~\cite{wen2010} by setting $B_m=0$. The solid curves are for
MSQM, while the dotted ones are for the corresponding nonmagnetized
SQM. The two groups of curves have apparently similar density
behavior. Obviously, however, the MSQM has lower energies than the
nonmagnetized SQM. To show the effect of different coupling
constants, we adopt three values of $g$.

\begin{figure}[ht]
\centering
\includegraphics[width=7cm,height=7cm]{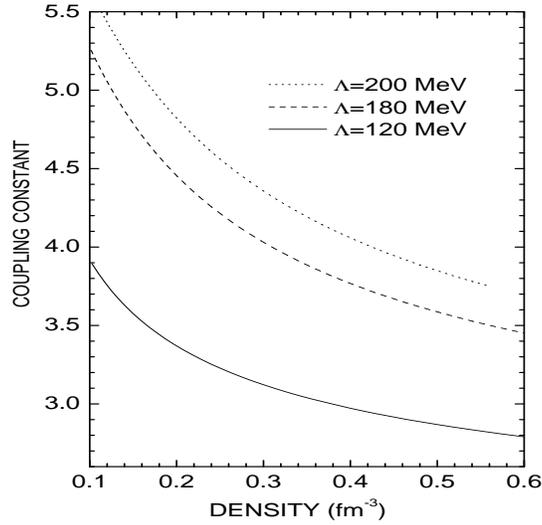}
\caption{ The running coupling constant $g$ versus the baryon number
density at different $\Lambda$ values with the magnetic field
B$_m=10^{17}$G. The upper lines correspond to larger values of
$\Lambda$.
        }
\label{coupl}
\end{figure}

In the quasiparticle model, the parameter $g$ stands for the
coupling strength and it is related to the strong interaction
coupling constant $\alpha_s$ by $g=\sqrt{4\pi \alpha_s}$. Therefore,
the g value has a large effect on the stability of
SQM~\cite{wen2011}. To satisfy the requirement of QCD asymptotic
freedom, the running property of the coupling parametrization should
be considered. In Fig. \ref{coupl}, we show the running coupling
constant as functions of the baryon number density $n_B$. The three
lines are obtained with different values of $\Lambda$. It is very
obvious from Fig.~\ref{coupl} that the running coupling $g$ is a
decreasing function of the density. With a bigger $\Lambda$ value,
the coupling $g$ is also bigger at any fixed density.

\begin{figure}[htb]
\centering
\includegraphics[width=7cm,height=7cm]{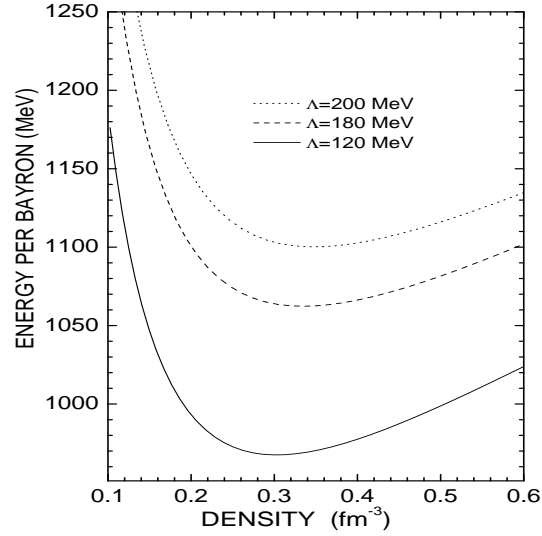}
\caption{The energy per baryon $E/n_B$ of stable MSQM versus the
number density at different $\Lambda$ values with the corresponding
critical magnetic field strength~B$_m^c$.
        }
\label{energ-nb}
\end{figure}

In Fig.~\ref{energ-nb}, we show the same quantities as in
Fig,~\ref{En} with the running coupling constant, respectively for
the two values of the different magnetic field $10^{17}$~G (dashed
lines) and $10^{18}$G (solid lines). It is clearly seen that the
energy per baryon increases with increasing the QCD scale parameter
$\Lambda$, i.e. SQM has a lower energy per baryon with smaller
$\Lambda$ value at a fixed strong magnetic field. This effect of the
QCD scale parameter is consistent with the constant coupling case in
Fig.~\ref{En}, because larger $\Lambda$ means bigger coupling as
indicated by Eq.~(\ref{g_run}).

\begin{figure}[htb]
\centering
\includegraphics[width=7cm,height=7cm]{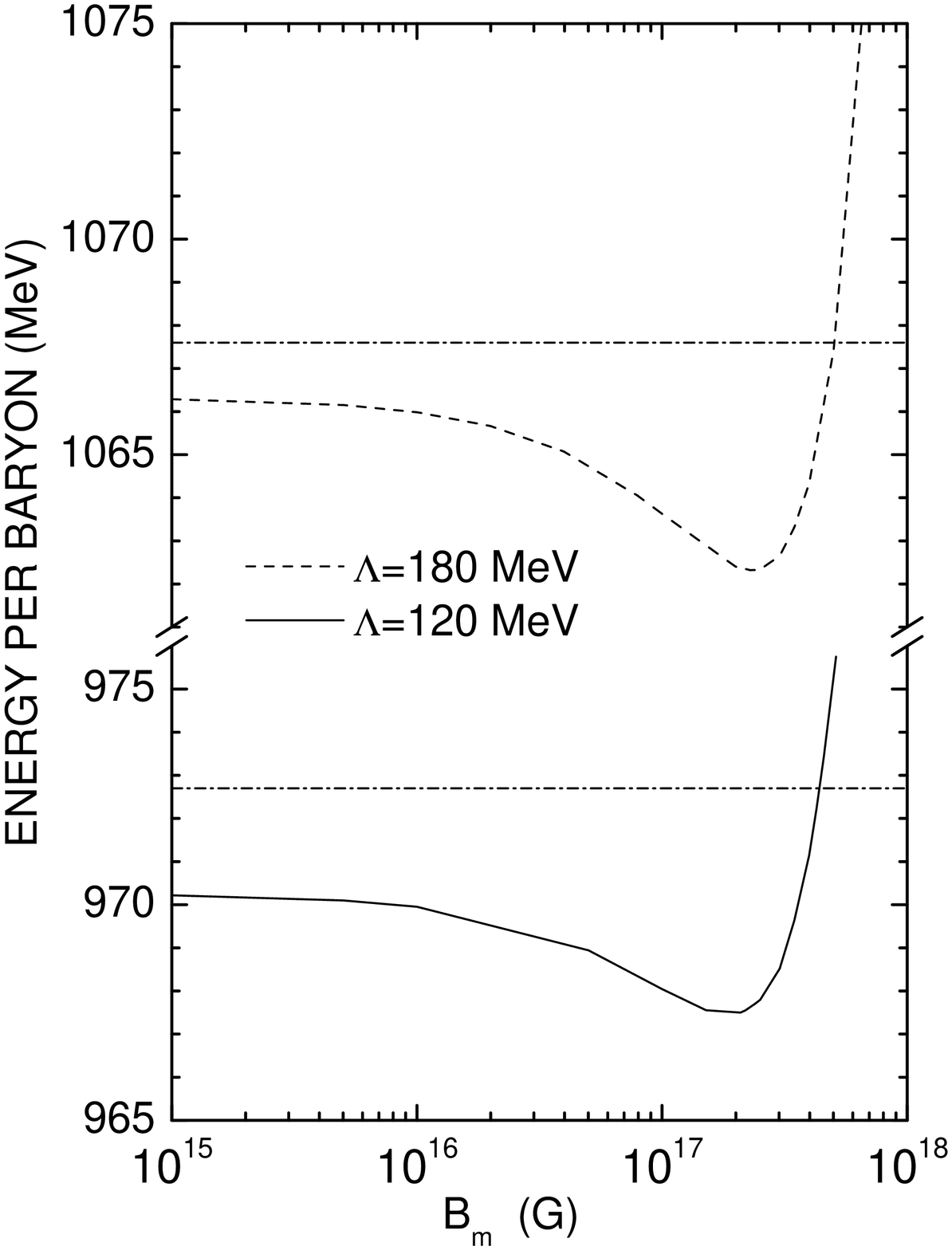}
\caption{The energy of magnetized strange quark matter varies with
the magnitude of the strong magnetic field for the fixed QCD scale
parameter $\Lambda=180$~MeV (the upper dashed curve) and
$\Lambda=120$~MeV (the lower solid curve) respectively. With
decreasing the magnetic field strength, the energy per baryon
approaches gradually to the value without a magnetic field indicated
by a horizontal dashed-dotted line.
        }
\label{lambdaBm}
\end{figure}

An obvious observation from Fig.~\ref{energ-nb} is that there is a
minimum energy per baryon for each pair of the parameters $\Lambda$
and $B_m$. In Fig.~\ref{lambdaBm}, therefore, we show how the
minimum energy of MSQM varies with the magnetic field strength. The
QCD scale parameter is taken to be 180~MeV (the upper dashed curve)
and 120~MeV (the lower solid curve) respectively. It is found on
each curve that there is another minimum value corresponding to a
critical magnetic field strength $B_m^c$. For the values of
$\Lambda=120$~MeV and $180$~MeV, the corresponding  $B_m^c$ equals
$2.15\times 10^{17}$~G and $2.34\times 10^{17}$~G respectively. When
the magnetic field strength is less than $B_m^c$, the minimum energy
per baryon decreases with increasing the strength of the magnetic
field. When the magnetic field strength exceeds $B_m^c$, or,
equivalently, when the magnetic energy scale approaches the QCD
scale, i.e., $\sqrt{eB_m}\sim 76.9$~MeV, the field energy itself
will have a considerable contribution to the energy of SQM and hence
the energy per baryon increases with the magnetic field strength. In
Fig.~\ref{energ-nb}, the magnetic field strength is taken to be the
corresponding critical value.

Because we study magnetized strange quark matter in the
"unpolarized" approximation, it is appropriate to estimate the
maximum magnetic field strength when such an approximation can be
reliable. To this end, in principle, we can investigate the
polarized quarks with spin up (+) and down (-) by introducing the
polarization parameter $\xi_i$ as \cite{Gonz08,Bordbar11}
\begin{eqnarray}\xi_i=\frac{n_i^{(+)}-n_i^{(-)}}{n_i^{(+)}+n_i^{(-)}}.
\end{eqnarray}
where $n_i^{(+)}$ and $n_i^{(-)}$ denote the number density of
spin-up and -down $i$-type quarks. For the sake of simplicity, we
assume a common polarization rate $\xi$ for $u$-, $d$-, and
$s$-quarks, i.e.,
$\xi_\mathrm{u}=\xi_\mathrm{d}=\xi_\mathrm{s}=\xi$. In
Sec.\ref{thermody}, the summation for fixed spin $s=+1$ or $s=-1$
should go over the principal quantum numbers $n$ instead of $\nu$.
The degeneracy factor ($2-\delta_{\nu 0}$) in Eqs.(\ref{eq:omega}),
(\ref{omega0}), (\ref{eq:dOdm}) and (\ref{eq:ni}) should be deleted
because the spin degeneracy disappears for polarized particles. The
polarization parameter $0\leq\xi\leq1$ will decrease with increasing
the number density. Assuming a larger value of the polarization
$\xi=0.6$, the energy is enlarged by $4.5\%$. In fact, even for very
larger magnetic field $B_m=5\times 10^{18}$G , the parameter $\xi$
remains in the range ($0.01\sim 0.02$) when the density $n_B>0.2$ fm
$^{-3}$ \cite{Bordbar11}. We do the numerical calculation and find
that the free energy per baryon will be enlarged by $0.8\%$ at
$\xi=0.1$. So the effect of the unpolarized approximation on the
discussion of the stability of SQM is very small, especially when
the magnetic strength is less than $10^{18}$G which is an estimated
maximum possible strength of the interior magnetic field.

\subsection{The effective bag function for magnetized SQM}

The effective bag function $B^*$ is generally used to represent the
vacuum energy density for dense QCD matter~\cite{Agui03}. Comparing
it with the standard statistical mechanics, one can recover the
thermodynamics consistency of system density- and/or
temperature-dependent Hamiltonian with the extra term $B^*$. The
meaning of $B^*$ plays an important role in studying properties of
quark matter. The interpretation of $B^*$ was first given by
Gorenstein and Yang in Ref.\cite{Goren1995}. In quasiparticle model,
because the dispersion relation is density- and/or
temperature-dependent, $B^*$ is regarded as the system energy in the
absence of quasiparticle excitations, which cannot be discarded from
the energy spectrum \cite{Gardim07}. In this sense, $B^*$ acts as
the bag energy or bag pressure through the application in bag-like
model. One can interpret the confinement mechanism considering $B^*$
as the difference of perturbative vacuum and physical vacuum.

\begin{figure}[ht]
\centering
\includegraphics[width=7cm,height=7cm]{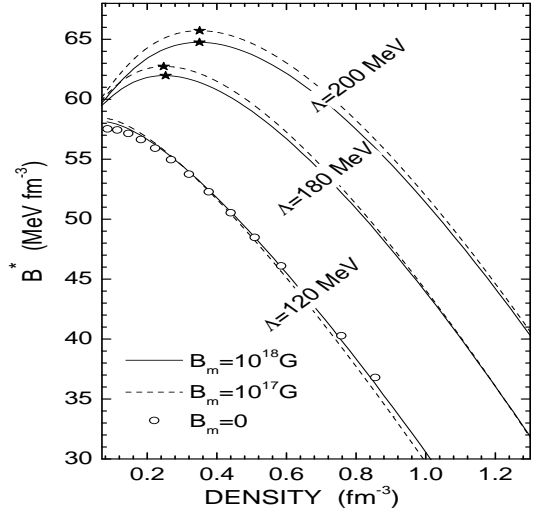}
\caption{The effective bag function $B^*$ for MSQM versus the baryon
number density at different $\Lambda$ values. $B_m=10^{17}$G and
$B_m=10^{18}$G are marked by dashed lines and solid lines,
respectively. Only for larger $\Lambda$, the $B^*$ has a maximum
value $B^*_\mathrm{max}$ indicated by asterisks at $2\sim 3$ times
nuclear saturation density.
        }
\label{Bstrun}
\end{figure}

In addition to the constant value $B_0$ of the bag model, the
expression of $B^*$ has been developed in several different forms.
Li, Bhalerao, and Bhaduri obtained the temperature-dependent bag
constant in the QCD sum-rule method~\cite{Li91}. Song obtained a
$\mu$- and $T$-dependent bag constant by incorporating one-loop
correction in imaginary time formulation of finite temperature field
theory~\cite{Song92},
\begin{equation}B^*(\mu,T)=B_0-[\frac{1}{162\pi^2}\mu^4+\frac{1}{9}\mu^2T^2+\frac{7\pi}{30}T^4].
\end{equation}
In the work of Burgio~\cite{Burgio02}, the Gaussian parametrization
of density dependence of $B^*$ is employed as,
\begin{equation}B^*(n_B)=B_\infty+(B_0-B_\infty)\exp(-\gamma(n_B/n_0)^2),
\end{equation}
where the parameters $B_\infty$, $\gamma$, and $n_0$ are given in
Ref.~\cite{Burgio02}. The effective bag constants in these previous
works are all monotonically decreasing functions of the density and
temperature \cite{Prasad04}. In our present work, the effective bag
function $B^*$ is associated with a magnetic field and consequently
has a different density behavior. We thus plot the effective bag
function $B^*$ versus the baryon number density with different
$\Lambda$ values in Fig.~\ref{Bstrun}. The dashed lines are for the
magnetic field strength $B_m=10^{17}$G, while the solid lines are
for a higher magnetic strength $B_m=10^{18}$G. The open circles
indicate nonmagnetized SQM. The numerical results show an important
property that the effective bag function $B^*$ remains decreasing
monotonously with increasing densities for smaller
$\Lambda=120$~MeV. But for larger value $\Lambda=180$ or $200$~MeV,
the bag function $B^*$ has a maximum value at about $2\sim 3$ times
the nuclear saturation density $0.16$ fm$^{-3}$. Generally, when the
QCD scale parameter is bigger than the critical value $126$ MeV, the
effective bag function is not a monotonic function and reaches a
maximum value $B^*_\mathrm{max}$ at the density range $0.3\sim 0.4$
fm$^{-3}$.

\begin{figure}[ht]
\centering
\includegraphics[width=7cm,height=7cm]{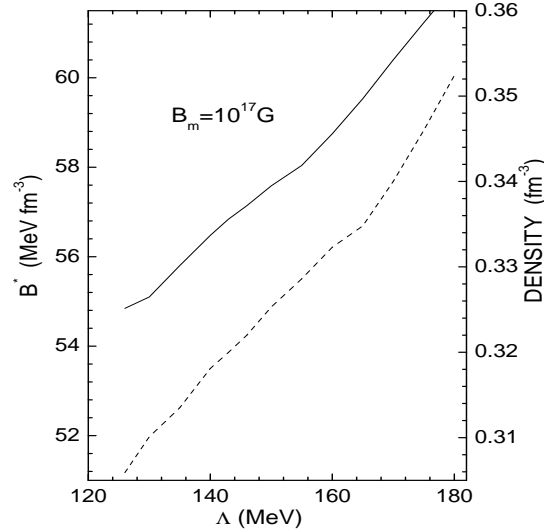}
\caption{$\Lambda$ dependence of the bag function $B^*$ (solid line)
and baryon number density (dashed line) corresponding to the zero
pressure. The $\Lambda$ should be larger than the critical value 126
MeV to produce a nonmonotonic behavior of $B^*$.
        }
\label{lambdanb}
\end{figure}

Since the QCD scale parameter $\Lambda$ plays a great role on the
effective bag function $B^*$, we plot the bag function $B^*$ of
stable SQM, i.e., $P=0$, versus $\Lambda$ on the left axis in
Fig,~\ref{lambdanb}. If one requires that the bag function $B^*$
should be a nonmonotonic decreasing function of the density, the
$\Lambda$ value should be bigger than the critical value $126$~MeV.
The corresponding baryon number density $n_B$ marked by a dashed
line on the right axis is also plotted. The bag function $B^*$ and
the baryon number density $n_B$ all increase with the QCD parameter
$\Lambda$.

\subsection{Mass-radius relation of magnetized strange quark stars}
\begin{figure}[ht]
\centering
\includegraphics[width=7cm,height=7cm]{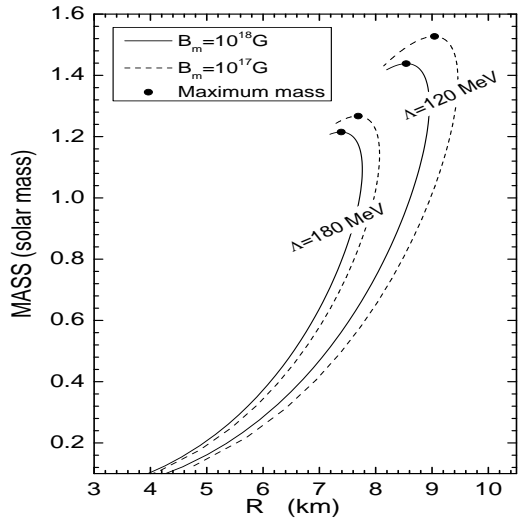}
\caption{The mass-radius relation of SQS at different $\Lambda$
values with different magnetic fields B$_m=10^{17}$G (dashed lines)
and B$_m=10^{18}$G (solid lines). The maximum masses on all curves
are marked by full dots.
        }
\label{MR}
\end{figure}

Strange quark stars, a family of compact stars consisting completely
of deconfined $u$, $d$, $s$ quarks, have attracted a lot of
researchers. The gravitational mass (M) and radius (R) of compact
stars are of special interest in astrophysics. The strange quark
stars were studied by many authors as self-bound stars different
from neutron stars. It is pointed out that the possible
configuration of compact stars, such as the strange hadrons,
hyperonic matter, and quark matter core, can soften the equations of
state of neutron stars \cite{Sahu2001,Shen2002,Burgio2002}. In this
section, we calculate the mass-radius relation of magnetized SQS
together with the effective quark mass scale. Using the EoS of MSQM
in the proceeding sections, we can obtain M and R by numerically
solving Tolman-Oppenheimer-Volkoff(TOV) equations when fixing a
central pressure $P_c$. Varying continuously the central pressure,
we can obtain a mass-radius relation $\textit{M(R})$ in Fig.
\ref{MR}. The stable branches of the curves must satisfy the
condition $dM/dP_c > 0$. In this way, we can find the maximum mass
along the same curve, which is denoted by full dots in
Fig.~\ref{MR}. Other solutions, on the left side of the maximum
mass, are unstable and collapsible.

It is seen from Fig.~\ref{MR} that the maximum mass is bigger with a
smaller $\Lambda$ value and an extremely large magnetic field.
However, it is still not as big as the recently observed maximum
mass of PSR J1614-2230 \cite{Demo2010}. This may mean that a simple
ordinary phase cannot explain the large mass. Some new phases, e.g.,
the superconductivity phase in dense matter
\cite{wilczek01,Madsen01,Peng06}, should be further studied in the
future.

\section{summary}
\label{Sec:conls}

We have extended the quark quasiparticle model to study the
properties of strange quark matter in a strong magnetic field at
finite density. The self-consistent thermodynamic treatment is
obtained through an additional bag function. The bag function
depends not only on the quark chemical potentials but also on the
magnetic field strength $B_m$. By comparison with the nonmagnetized
quark matter, we find that the magnetic field can enhance the
stability of SQM when the magnetic field strength is lower than a
critical value of the order $10^{17}$~G. But when the magnitude of
the magnetic field is larger than the critical value $B_m^c$, the
magnetic energy will have a considerable contribution to the energy
of SQM. So the energy per baryon of MSQM increases with increasing
the field strength. Because the quark masses depend on the
corresponding chemical potential, an additional effective bag
function, which depends not only on the chemical potentials but also
on the magnetic field strength, appears in both the energy density
and pressure. The effective bag function has a maximum at about
$2\sim 3$ times the saturation density when the QCD scale parameter
is larger than $126$~MeV. Although an unpolarized approximation is
assumed, we find the energy per baryon would increase by $0.8\%$ for
the usual polarization parameter when $n_B>0.2$~fm$^{-3}$.

On application of the new equation of state of the magnetized
strange quark matter in ordinary phase to calculate the mass-radius
relation of a quark star, it is found that the maximum mass does not
explain the the newly observed maximum mass of about two times the
solar mass. This means that other phases, e.g. superconductivity
and/or mixed phases, might be necessary to explain the new
astronomic observations, and further studies are needed.

\begin{acknowledgments}
The work is supported by the National Natural Science Foundation of
China (11005071, 11135011, and 11045006) and the Shanxi Provincial
Natural Science Foundation (2011011001). X.J.W. specially thanks
Professor J.J. Liang at Shanxi University for helpful discussions.
\end{acknowledgments}

\end{document}